\documentclass{amsart}
\usepackage{amsmath,graphicx}
\usepackage{amssymb}
\usepackage[normalem]{ulem}
\usepackage{color}

\begin{document}

\newcommand\re[1]{(\ref{#1})}
\newcommand{\pd}[2]{\frac{\partial #1}{\partial #2}}
\newcommand{\pdt}[3]{\frac{\partial^2 #1}{\partial #2 \partial #3}}
\newcommand{\mg}[1]{\partial_{#1}}
\newcommand{\F}[2]{F^{#1}_{\;#2}}
\renewcommand{\labelitemi}{--}

\title{Universality in heat conduction theory -- \\
weakly nonlocal thermodynamics }
\author{P. V\'an$^{1,2,3}$ and T. F\"ul\"op$^{1,2,3}$}
\address{$^1$Department of Theoretical Physics\\
         Wigner RCP, RMKI, Budapest, Hungary and \\
$^2$Department of Energy Engineering\\
        Budapest University of Technology and Economics, Hungary\\
$^3$Montavid Thermodynamic Research Group}

\pagestyle{plain}
\markright{Universal Heat conduction}

\date{\today}

\begin{abstract}
A linear irreversible thermodynamic framework of heat conduction in rigid conductors is introduced. The deviation from local equilibrium is characterized by a single internal variable and a current multiplier. A   general constitutive evolution equation of the current density of the internal energy is derived by introducing linear relationship between the thermodynamic forces and fluxes. The well-known Fourier, Maxwell-Cattaneo-Vernotte, Guyer-Krumhansl,  Jeffreys type and Green-Naghdi type equations of heat conduction are obtained as special cases. The universal character of the approach is demonstrated by two examples.
Solutions illustrating the properties of the equation with jump boundary conditions are given. \end{abstract}

\maketitle

\section{Introduction}

The increasing importance of micro- and nanotechnology initiated an intensive research in heat conduction \cite{CahEta03a,Zha07b}. Experiments show deviations from the classical Fourier theory \cite{KimEta01a,FujEta05a} and there are several theoretical developments to understand the nature of the deviations   \cite{CimEta09a,CimEta10a2,Cim09a,Tzo11a,GrmEta11a,BarSte07a,BarSte08a}.   

The starting point of all these approaches is the balance of internal energy
\begin{equation}
 \rho \dot e+ \partial^i q^i = 0, \label{ebal} 
\end{equation}
where $\rho$ is the density, $e$  is the specific internal energy, and $q^i$  is the conductive part of the current density of internal energy, i.e., the heat flow. 
{We are dealing with rigid heat conductors,  
}the dot denotes the partial time derivative,  $\partial^{i}$ is the space derivative of the corresponding physical quantity, and index notation with the Einstein summation convention is applied. In the phenomenological generalizations  of Fourier's law of classical heat conduction, the constitutive equation  \re{Four}  is modified by additional terms. The most important modifications are the following:
\begin{eqnarray}
q^ i &=& -\lambda \partial^i T,\label{Four}\\ 
\tau \dot q^i+q^ i &=& -\lambda \partial^i T,\label{CatVer}\\ 
\tau \dot q^i+q^ i &=& -\lambda \partial^i T +a_1\partial^{ij}q^j+a_2\partial^{jj}q^i,\label{GuyKru}\\ \tau \dot q^i+q^ i &=& -\lambda \partial^i T+b_{2}\partial^i\dot T,\label{Jef}\\
\tau \dot q^i \qquad  &=& -\lambda \partial^i T+ a_2 \partial^{jj}q^i .\label{GreNag}
\end{eqnarray}

Here,  \re{Four} is the classical Fourier's law \cite{Fou55b},  \re{CatVer} is the Maxwell-Cattaneo-Vernotte (MCV) equation \cite{Max867a,Cat48a,Ver58a1}, Eq. \re{GuyKru} is the Guyer-Krumhansl (GK) equation \cite{GuyKru66a1},   \re{Jef} is known as the Jeffreys type or lagging heat equation \cite{JosPre89a}, and  \re{GreNag} leads to the Green-Naghdi (GN) equation \cite{GreNag91a}. 
The heat conduction coefficient is denoted by \(\lambda\), the relaxation time by \(\tau\), and \(a_1\), \(a_2\)  and \(b_{2}\) are further material parameters. 

The origin---the motivation and derivation---of these equations is manifold. Kinetic theory offers several different approaches for Fourier's law and for MCV equation. For instance, moment series expansion of the Boltzmann equation results in  Fourier's law in first order and in the MCV equation in second order \cite{JouAta92b,MulRug98b}.  Also, the GK equation  was originally derived with the help of special collision terms in the Boltzmann equation characterizing phonon-lattice interaction  \cite{GuyKru66a1,GuyKru66a2}.  

The phenomenological theories are also diverse regarding the origin of the above equations \cite{JosPre89a,JosPre90a,Cim09a}. Fourier's  law is the consequence of non-negative entropy production in classical irreversible thermodynamics \cite{GroMaz62b}, and the MCV equation is the result of the deviation from local equilibrium, characterized by an internal variable \cite{FabMor03a}. The heat flow is the well-established candidate of that internal variable to obtain compatibility with kinetic theory  \cite{Gya77a,JouAta92b,MulRug98b}. 

The origin of weakly nonlocal extensions is a question  in a phenomenological approach, too.  There are indications that the  Guyer-Krumhansl equation is connected to nonclassical entropy current density \cite{Van01a2,CimVan05a,CiaAta07a}. The lagging heat equation \re{Jef} was suggested according to an analogy to  rheological models and its relation to thermodynamics is not clear \cite{Cim09a}.  One may obtain it by diverse simple mechanisms of temperature equilibration, like heat flow splitting  \cite{TamZho98a}, or two-step relaxation \cite{AniEta74a,FujEta84a}. The Jeffreys type heat conduction is free from several problematic aspects of the MCV equation \cite{BriZha09a}. At last, the theory of Green and Naghdi  is peculiar. They have  introduced a special scalar internal variable, the time derivative of which is the temperature, and have analysed the thermodynamic consequences of the deviations from local equilibrium in a nonstandard  way \cite{GreNag91a}. They have obtained a model of heat conduction that cannot be reduced easily to Fourier's law, and which may result in  zero entropy production: heat conduction without dissipation.

We can see from the previous concise  and not complete survey that the derivation and the motivation of the weakly nonlocal generalizations of Fourier's  theory are manifold. Most of them are considered valid models of physical phenomena if their microscopic derivation provides a clear mechanism of the modifications. In some important cases---like the Jeffreys type or lagging heat equation---the role of the second law is not clarified. All kinds of thermodynamically consistent  justifications introduce a deviation from the local equilibrium entropy density. In case of nonlocal extensions, a deviation from the classical form the entropy current density is also considered. 

In this paper, we derive a unified model of the previous constitutive relations with the help of irreversible thermodynamics.  We need only two simple and general assumptions: 
\begin{itemize}
\item there is a deviation from the equilibrium state, conveniently and generally characterized by a vectorial  \textit{internal variable} \cite{Ver97b,MauMus94a1,MauMus94a2}, and 
\item there is a deviation from the classical form of the entropy current, conveniently and generally characterized by an arbitrary function that we will call 
\textit{current multiplier} \cite{CimVan05a}.
\end{itemize}
In a linear approximation, both the internal variable and the current multiplier can be completely eliminated and one obtains a general constitutive equation for the heat conduction. We will see that the restrictions from the second law of thermodynamics are nontrivial and reduce the number of independent coefficients. 
{
In this paper, we apply the simple heuristic constitutive theory of irreversible thermodynamics in order to consider the restrictions of the second law. A detailed second law analysis of rigid heat conductors  with the help of Liu procedure in \cite{CimVan05a} shows  that current multipliers require second order weak nonlocality.
}

There is an important advantage of a phenomenological thermodynamic approach based only on general assumptions. As long as the general conditions regarding the deviation from the local equilibrium are fulfilled by any microscopic or mesoscopic assumption regarding the mechanism and the structure of the deviation, the consequences will be the same: this is the universality of nonequilibrium thermodynamics. To demonstrate this property, we show, that if the deviation from Fourier's law is characterised by a gradient of a scalar field, then the model equations reduce to the parabolic two-step model,  and that, if the deviation from  Fourier's law is characterised by a general vector field, then the vector field should fulfil an MCV equation in order to satisfy the thermodynamic constraints.

\section{Irreversible thermodynamics of heat conduction}

\subsection{The entropy production}

For modeling phenomena beyond local equilibrium, we introduce a single vectorial internal variable denoted by \(\xi^{i}\), and through this paper we consider {rigid }isotropic materials. The  second law is given in the following form:
\begin{equation}
\rho \dot s + \partial^i J^i = \sigma \geq 0, 
\label{entbal}\end{equation}     
where \(s\) is the specific entropy, \(J^i\) is the conductive current density of  entropy and  \(\sigma\) is the entropy production. Then we introduce two basic constitutive hypotheses, which will be crucial in the following:\ 

\begin{enumerate}
\item We assume that nonequilibrium entropy depends on the internal variable \(\xi^i\)   quadratically:
 \begin{equation}
s(e,\xi^{i}) = \hat s \left( e \right)- \frac{m}{2} \xi^2.
\label{entrint}\end{equation}
  
 Here, \(m \) is a scalar material coefficient, sometimes called thermodynamic inductivity \cite{Gya77a}. The quadratic dependence, with \(m=m(e,\xi^{i})\) can be justified by the Morse lemma and the requirement of entropy maximum at the nonequilibrium part of the basic state space, spanned by \(\xi^i\) \cite{Ver97b}. $m\geq 0$ because of the concavity of entropy. If \(m\) is constant then we obtain the following partial derivatives:
\begin{equation}
\left. \frac{\partial s}{\partial e}\right |_{\xi^i} = \frac{1}{T}, \qquad
\left. \frac{\partial s}{\partial \xi^i}\right |_e = -m\xi^i,
\label{intensives}\end{equation}
where \(T\) is the equilibrium temperature.  The thermodynamic relations are conveniently expressed by the following Gibbs relation:
\begin{equation}
de=Tds+m\xi^iTd\xi^i.
\label{Gibbsrel}\end{equation}

{In what follows, we make the assumption \(m=const.\), which is   convenient in the following calculations, although it restricts the applicability of the theory. Then, the temperature is the function of the internal energy only and does not depend on the internal variable, as one can see from the first equation of \re{intensives}. }

\item Our second assumption requires the generalization of the classical conductive current density of the entropy. We demand that, without energy current, there is no entropy current in pure heat conduction. Therefore, in the following we assume that, instead of the classical \( J^i = q^i/T\) form, the current density of the entropy is  
\begin{equation}
J^i = B^{ij}q^j,
\end{equation}\label{Nyiriec}
\!\!\! where   \(B^{ij}\) is a \emph{constitutive function} that will be constrained by the second law.  This assumption was first introduced by Ny\'iri \cite{Nyi91a1}, and applied for heat conduction in \cite{Van01a2}, where the generalized multiplier of the current density of the internal energy was called current intensity factor. 
The recent terminology ('current multiplier') was suggested in \cite{CimVan05a}.  
\end{enumerate}

The above two conditions lead to the following form of the entropy production:
\begin{eqnarray}
\rho\dot s + \partial^i J^i &=& -\frac{1}{T} \partial^i q^i - 
    {\rho}m\xi^i \dot \xi^i + \partial^i(B^{ij}q^j) = \nonumber\\
 && \left(\partial^j B^{ij}\right)q^i +
 \partial^i q^j\left(B^{ij} -\frac{1}{T}\delta^{ij} \right) - 
 {\rho m} \xi^i \dot \xi^i \geq 0.
\label{rhsepr}\end{eqnarray}

{This formula is a quadratic expression of thermodynamic fluxes and forces, where the following characterisation is applied: 
\renewcommand{\arraystretch}{1.5}
\begin{center}
\begin{tabular}{c|c|c|c}
       & Classical thermal & Extended thermal & Internal \\ \hline
Fluxes & $ q^i$ & 
    $B^{ij} -\frac{1}{T}\delta^{ij} $ & 
    ${\rho m}  \dot \xi^i$\\ \hline
Forces &$\partial^j B^{ij}$ &
    $\partial^i q^j$ &
    $- \xi^i $\\
    \end{tabular}\\
\vskip .21cm
{Table 1. Thermodynamic fluxes and forces}\end{center}
The heuristic method of classical irreversible thermodynamics introduces a functional relationship between these quantities, assuming that the fluxes depend on the forces. Then in simple cases and  for twice differentiable functions  the mean value theorem of Lagrange ensures the validity of linear relations \cite{Van03a}. However, in our case the bilinear expression above does not permit such a simple characterization, the natural choice of the constitutive state space spanned by the internal energy \(e\) and the vectorial internal variable \(\xi^i\) and their gradients is seemingly not without problems. However, in \cite{CimVan05a} it was shown that, in case of a second order weakly nonlocal state space, the above choice of the entropy flux in the form of \re{Nyiriec} does not lead to a contradiction  and, considering also that composite functions can appear in constitutive relations (see e.g. \cite{Van08a}), the consequent simple heuristic exploitation can be corroborated. }

\subsection{Linear relations} {Now we introduce  such linear relationships that can be easily transformed into a convenient form. In isotropic continua the entropy production is an isotropic function of the forces, therefore, the} most general linear relationship \cite{Smi71a} between them introduces seven material parameters:
\begin{eqnarray}
q^i &=& l_1 \partial^j B^{ij} - l_{12} \xi^i, \label{o1}\\
m\rho \dot \xi^i &=& l_{21} \partial^j B^{ij} -  l_{2} \xi^i, \label{o2}\\
B^{ij}-\frac{1}{T}\delta^{ij} &=& k_1 \partial^i q^j + k_2 \partial^j q^i + 
    k_3 \partial^k q^k \delta^{ij}. 
\label{o3}\end{eqnarray}
  Here, $l_1$, $l_{12}$, $l_{21}$, $l_{2}$, $k_{1}$, $k_{2}$, $k_{3}$ are the isotropic scalar conductivity coefficients, and $\delta^{ij}$  is the Kronecker symbol. The non-negative entropy production requires the following inequalities for the material parameters: 
\begin{gather}
l_{1}\geq 0, \quad l_2\geq 0, \quad k_1\geq 0, \quad  k_2\geq 0, \quad  k_3\geq 0, \nonumber\\  L= l_1l_2 - \frac{1}{4}(l_{12}+l_{21})^2 \geq 0.
\label{poz}\end{gather}

We do not assume {Onsager or Casimir type} reciprocal relations for the vectorial thermodynamic interactions characterized by the last two terms of the entropy production. Our internal variable may be a function of several of microscopic variables of different nature regarding time reversibility, {some \(\alpha \) type and others \(\beta\).} This general approach was shown to be fruitful in formulating the thermodynamic framework of generalized continuum mechanics \cite{VanAta08a,BerEta10a}.

We can eliminate the current multiplier from \re{o1} and \re{o2} with the help of \re{o3}. Moreover, after some simple steps, we can eliminate the internal variable from Eq.\re{o1} and Eq.\re{o2}, too. If \(l_2 \neq 0\) and \(m,l_{1},l_{12},k_1,k_2,k_3\) are constants then we obtain the following constitutive relationship of the derivatives of temperature and of energy current density: 
\begin{multline}
\tau\frac{\partial}{\partial t} q^i +q^i =\\
 \lambda_1 \partial^i \frac{1}{T} +
     \lambda_2 \frac{\partial}{\partial t}\left(\partial^i \frac{1}{T}\right) + 
     a_1 \partial^{ij} q^j + a_2 \partial^{jj} q^i
     + b_1  \frac{\partial}{\partial t}(\partial^{ij}q^j)  + 
     b_2  \frac{\partial}{\partial t}(\partial^{jj}q^i). 
\label{geneq}\end{multline}

Here, we have denoted the {partial} time derivative by $\partial/\partial t$ and introduced the shorthands
\begin{gather}
\tau=\frac{m\rho}{l_2}, \qquad 
\lambda_1= l_1-\frac{l_{12}l_{21}}{l_2}, \qquad
\lambda_2= m\rho\frac{l_1}{l_2}, \nonumber\\
a_1= \lambda_1(k_1+k_3), \qquad
a_2= \lambda_1k_2, \label{ujpar}\\
b_1= \lambda_2(k_1+k_3), \qquad
b_2= \lambda_2k_2.
\nonumber\end{gather}
 We can see that  \re{geneq} contains only five independent material parameter, as \(k_1\) and \(k_3\),  \(l_{12}\) and \(l_{21}\) cannot be distinguished. Every coefficient in \re{ujpar} is positive  according to \re{poz}. The   heat conduction coefficient deserves a special attention because \begin{equation}
l_1l_2 - l_{12}l_{21} = l_1l_2 - l_s^2 + l_{a}^2\geq 0, 
\label{lap}\end{equation}
where \(l_{s}= (l_{12}+l_{21})/2\) and \(l_a= (l_{12}-l_{21})/2\) are the symmetric and anti symmetric parts of the matrix in \re{o1}-\re{o2}.
As a consequence if \(\lambda_1 =0\), then \(\lambda_2 =0\) follows.

\subsection{Special cases} We derive the following important special cases:

\begin{enumerate}
\item\textit{Fourier.}  If  \(k_{1}=k_{2}=k_{3}=0\) and \(l_{12}=0,\) then directly from  \re{o1}-\re{o3} we obtain Fourier's law,  in the following form:
\begin{equation}
q^ i = \lambda_1 \partial_i \frac{1}{T} =-\lambda \partial_i T, 
\label{TFour}\end{equation}
where \(\lambda=\lambda_1/T^2 = l_1/T^2\) is the Fourier heat conduction coefficient. This is not apparent from \re{geneq} because  \re{ujpar} and \re{poz} exclude the straightforward  choice  \(\tau = 0\), \(\lambda_2=0\), \(a_1=a_2=0\), \(b_1=b_2=0\).

\item \emph{Maxwell-Cattaneo-Vernotte}. It is frequently mentioned \cite{JouAta92b} that extended irreversible thermodynamics, and the MCV equation, arises as a special choice of a vectorial internal variable as the conductive current density of the internal energy \(\xi^i = q^i\) \cite{MauMus94a1,Ott05b}. {However, in our framework, the choice \(\xi^i= {q}^i\) does not automatically lead to the MCV equation. } In fact, \re{geneq} shows that the choice of  \(\lambda_2=0\), \(a_1=a_2=0\), \(b_1=b_2=0\), leads to equation \mbox{\re{CatVer}\!.} Thence  \(l_{1}\)=0, and therefore, the internal variable is proportional to the heat flow according to \re{o1}. \( \lambda_1=l_{a}^2/l_2\) because of the last inequality of \re{poz}. {Since in \re{o1} the first term of the right hand side is responsible for the classical Fourier type heat conduction, }we may say he MCV equation is obtained if heat conduction is dominated by a Casimir type antisymmetric  cross effect.

\item\textit{Jeffreys type.} If  \(a_1=a_2=0\), \(b_1=b_2=0\), we obtain the thermodynamic version of the Jeffreys type equation in the following form:
\begin{equation}
\tau\frac{\partial}{\partial t} q^i +q^i =\\
 \lambda_1 \partial^i \frac{1}{T} +
     \lambda_2 \frac{\partial}{\partial t}\left(\partial^i \frac{1}{T}\right). 
\label{TJeff}
\end{equation}
If \(l_{1} \neq 0\) then \(\lambda_2 \neq 0\) follows, and the MCV equation is completed to a Jeffreys type equation.
The emerging nonlinearity cannot be circumvented by assuming temperature dependent coefficients. 

\item\textit{Guyer-Krumhansl.} If \(\lambda_2=0\), \(b_1=b_2=0\), and \(\lambda_1 = \lambda T^2\), then the GK equation \re{GuyKru} is obtained. {As \(l_1=0\) is the simplest condition for \(\lambda_2=0\), we see that the} GK equation requires a Casimir type coupling of the terms in \re{o1}--\re{o2}. 

\item\textit{General Green-Naghdi type.} A GN  type equation can be obtained if \(l_{2}=0\). In this case, the  Casimir type reciprocity 
{ \(l_a=l_{12}=-l_{21}\)}
is the consequence of the last inequality of \re{poz}, and we obtain
\begin{multline}
\hat\tau\frac{\partial}{\partial t} q^i =
\hat \lambda_1 \partial^i \frac{1}{T} +
     \hat\lambda_2 \frac{\partial}{\partial t}\left(\partial^i \frac{1}{T}\right) \\ 
     +\hat a_1 \partial^{ij} q^j + \hat a_2 \partial^{jj} q^i
     + \hat b_1  \frac{\partial}{\partial t}(\partial^{ij}q^j)  + 
     \hat b_2  \frac{\partial}{\partial t}(\partial^{jj}q^i). 
\label{gGNeq}\end{multline}
Here, we have introduced similar notation of the coefficients as in \re{geneq}, but they are combinations of the thermodynamic parameters different from that was given in \re{ujpar}:
{\begin{gather}
\hat\tau={m\rho}, \qquad 
\hat\lambda_1=l_{a}^2, \qquad
\hat\lambda_2= m\rho{l_1}, \nonumber\\
\hat a_1= \hat \lambda_1(k_1+k_3), \qquad
\hat a_2= \hat \lambda_1k_2, \label{GNpar} \\
\hat b_1= \hat \lambda_2(k_1+k_3), \qquad
\hat b_2= \hat \lambda_2k_2.
\nonumber\end{gather}}

\item\textit{Green-Naghdi type.} A pure GN type equation requires  \(l_{2}=0\), \(l_1=0\), and the Casimir type reciprocity \(l_a=l_{12}=-l_{21}\). Then we obtain
\begin{equation}
\hat\tau\frac{\partial}{\partial t} q^i = \hat\lambda_1 \partial^i \frac{1}{T} +
         \hat a_1 \partial^{ij} q^j + \hat a_2 \partial^{jj} q^i.
\label{GN}\end{equation}
 The GN type equation may be nondissipative with zero entropy production when the entropy current density is classical, i.e. \(k_1=k_2=k_3=0\). 
\end{enumerate}

{An observation common to all the above cases is} that heat conduction coefficient \(\lambda_1\) is always nonnegative.  

\section{Macroscopic universality}  

In order to obtain an impression of the specific mechanisms that may lead to weakly nonlocal deviation from the classical entropy current, we consider here a simple form of the GK equation, where, in addition to the conditions \(\lambda_2=0\), \(b_1=b_2=0\), we introduce  \(k_{1}=k_2=0\). In this case, \re{geneq} simplifies to 
\begin{equation}
\tau \dot q^i +q^i = \lambda_1 \partial^i B,
\label{sGK}\end{equation}
with \(B={1}/{T}+k_3 \partial^kq^k\), \(\tau = m\rho/l_2\) and \(\lambda_1 = l_{a}^{2}/l_2\).

\subsection{Two heat fluxes} As a first example, let us suppose that the deviation from Fourier's heat conduction is characterized by a vector field \(\hat q^i\):
\begin{equation}
  \hat q^i:= q^i + \lambda_{S} \partial^i T,
\label{SGK}\end{equation}
with \(\lambda_S\) being a constant parameter. We may substitute \(q^{i}\) into the simplified GK equation \re{sGK} and obtain the following condition for the unknown vector field:
\begin{equation}
\tau \dot {\hat q}^i +\hat q^i = (\lambda_1-\lambda_S T^2)\partial^i\frac{1}{T} + (\tau\lambda_S-\lambda_1 k_3 \rho c)\partial^i \dot T.
\end{equation}
Here we applied \re{ebal}, together with the equation of state $e=cT$ with constant specific heat \(c\). Hence, with the choice of \(k_3 = \tau\lambda_S/(\lambda_1\rho c) \), our vector field \(\hat q^i\) fulfils an MCV equation with the \(\lambda = \lambda_S -\lambda_{1}/T^2\) Fourier heat conduction coefficient. Therefore, the total heat flow \(q^i\) is divided into two parts: \(q^i-\hat q^i\) fulfils Fourier's law and \(\hat q^i\) an MCV equation. Apparently, there are two different channels of heat conduction characterized by a Fourier's law and an MCV equation, respectively \cite{TamZho98a}.
\subsection{Two temperatures} We can introduce a different characterization of the deviation from Fourier's law by a scalar field \(T_{2}\) in the following way:
\begin{equation}
\beta \partial^i T_2 = q^i + \lambda_{T} \partial^i T,
\label{TT}\end{equation}
where \(\lambda_T\) and \(\beta\) are constant coefficients. Substituting \(q^{i}\) to the simplified GK equation \re{sGK} again, one obtains the following condition:\begin{equation}
\partial^i\left(\beta(\tau \dot T_2+T_2-T) \right) = 
\left(\lambda_1+(\lambda_T-\beta)T^2\right)\partial^i\frac{1}{T} + (\tau\lambda_T-\lambda_1 k_3 \rho c)\partial^i \dot T.
\end{equation}
Therefore, with the choice of \(k_3 = \tau\lambda_T/(\lambda_1\rho c)\),  \(\lambda_1=(\beta-\lambda_T)T^2\) one can find the condition that our scalar field \(T_2\) fulfils a heat exchange equation. Apparently, there are two different temperatures that equilibrate by heat exchange. This is a well-known mechanism of heat conduction in complex materials  where one of the components can be thermally excited independently of the other, e.g. in case of metals with different electron and lattice temperatures \cite{AniEta74a,FujEta84a}.

Therefore we can see that both heat splitting and heat conduction of two-temperature materials are special cases of weakly nonlocal heat conduction and our approach with a general internal variable can be well interpreted in terms of the general weakly nonlocal heat conduction law \re{geneq}. The current multiplier \(B^{ij} = B \delta^{ij}\) is connected to the coefficients of the specific relaxation mechanism. 
These two examples show the real advantage of our weakly nonlocal generalization of heat conduction: it is based on minimal and general assumptions regarding the current density of the entropy and therefore the results are \emph{robust} and \emph{universal}. 

{A straightforward  microscopic interpretation of the internal variable could be a moment of the phonon distribution function. That view is suggested by the quadratic dependence of the entropy density of \(\xi\). However,  a detailed kinetic interpretation of current multipliers in the generalized entropy flux is an open question in the light of the above mentioned macroscopic ones.   } \

\section{Solutions of the generalized heat conduction equation}

In this section we solve a special form of the above equations for a one dimensional space variable \(x\).  Assuming a constant specific heat,  \(e=cT\) we can eliminate the heat flux from Eq.\re{geneq} with the help of the energy balance and obtain
\begin{equation}
\tau \frac{\partial^{2}T}{\partial t^2} + \frac{\partial T}{\partial t} =
   -\frac{\lambda_1}{\rho c} \frac{\partial^{2}}{\partial x^2} \frac{1}{T} 
   - \frac{\lambda_2}{\rho c} \frac{\partial^{3}}{\partial x^2\partial t}\frac{1}{T}+
   \lambda_1 k\frac{\partial^{3}T}{\partial x^2\partial t} + 
   \lambda_2k \frac{\partial^{4}T}{\partial x^2\partial t^2}
\label{hceq}\end{equation}
where  \(k=k_1+k_2+k_{3}\). 
We can see that, due to the appearance of the reciprocal temperature, the equation is nonlinear and the Jeffreys type and GK equations do not coincide. In this equation, we have introduced constant conduction coefficients based on \re{o1}--\re{o3} as it is natural in a thermodynamic approach. We can observe that, in this case, the Fourier heat conduction coefficient is temperature dependent,  \(\lambda_F=\lambda_1/T^2 \neq const.\) .

In the following, we show some characteristic solutions of the differential equation \re{hceq} in case of jump boundary conditions. Initially we assume that a rod with length $L$ has a fixed temperature   \(T(0,x)=T_i\) and also \(\partial_tT(0,x)=0\). The far end of the rod is kept at this initial temperature, therefore, \(T(t,L)=T_i\). The other boundary has the same temperature at the beginning, but after a short time the temperature jumps to a higher value \(T_{f}\). For the sake of calculation convenience, we approximate the jump with the following smooth function: \(T(t,0) = T_{i}+(T_{f}-T_i)(1-e^{- t/t_{\mathrm{jump}}})^{2}\). 

The used parameters are those of the NaF crystal experiments of Jackson and Walker \cite{JacWal71a,BarSte05a}.  Therefore the specific heat is \(  c=2.774\mathrm{W/Kkg}\), the density \(\rho =2866\mathrm{kg/m^{3}}\), the Fourier heat conduction coefficient  \(\lambda_F=20500\mathrm{W/mK}\), therefore,  \(\lambda_1=4.61\times 10^6 \mathrm{WK/m}\), the length of the rod is \(L=8.2 \mathrm{\mu m}\), and the parameters of the initial condition are \(T_i=15\mathrm{K},\ T_{i}-T_f=0.1\mathrm{K},\ t_{\mathrm{jump}}=0.5\mathrm{\mu s}\). 

\begin{figure}
\begin{center}
       \includegraphics[width=0.7\textwidth]{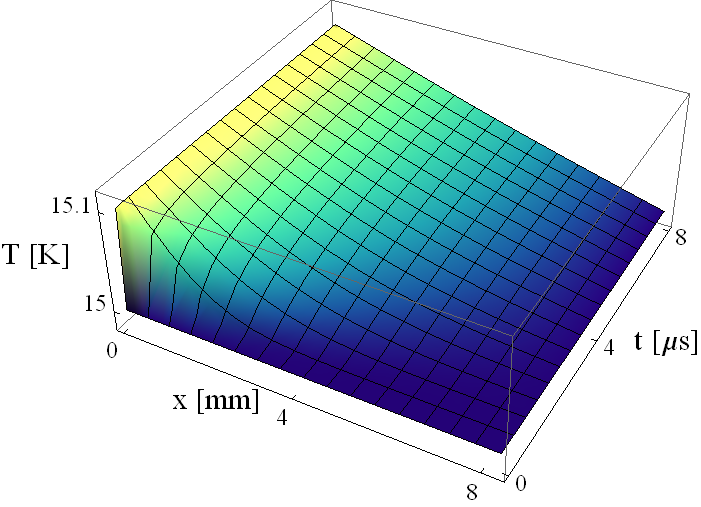}
\end{center}
\caption{ \label{Fourierfig}
Fourier equation, \(\tau = 0\mu s\).}
\end{figure}

Figure \ref{Fourierfig} shows the solution of the Fourier equation.   The linear stationary solution \(T(x)= T_{i}+T_{f}x\) is reached at the end. 

\begin{figure}
\begin{center}
        \includegraphics[width=0.7\textwidth]{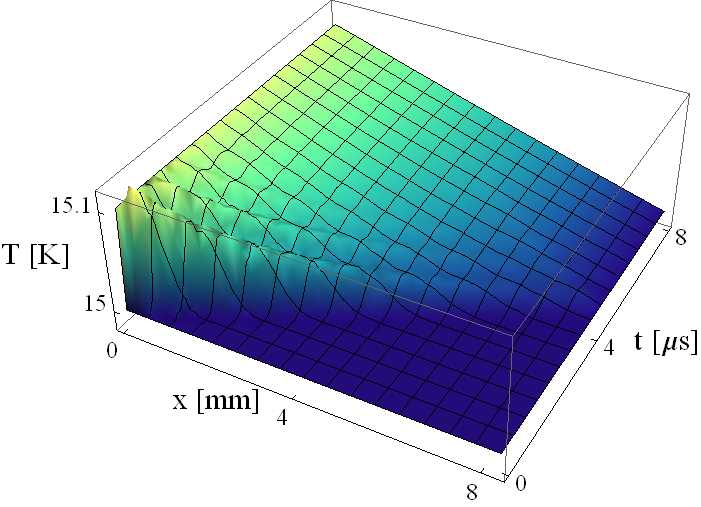}
\end{center}
\caption{ \label{MCVfig}
MCV equation,  \(\tau = 6.8 \mu s\).}
\end{figure}

\begin{figure}
\begin{center}
       \includegraphics[width=0.7\textwidth]{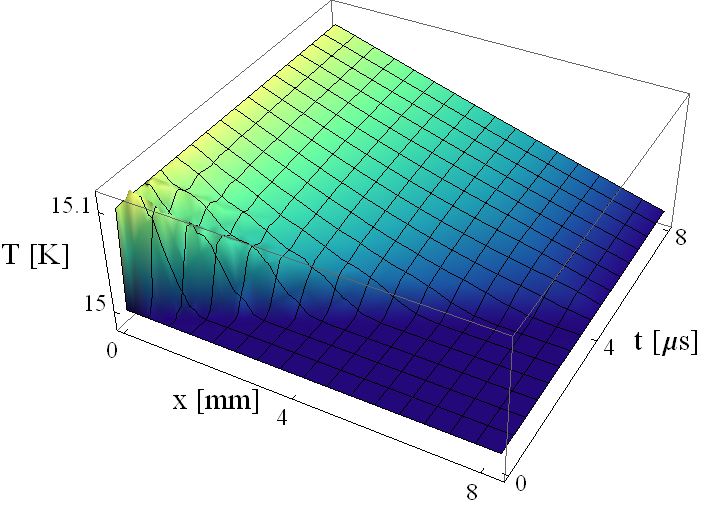}
\end{center}
\caption{ \label{Jeffreysfig}
Jeffreys type equation,  \(\tau = 6.8 \mu s\),  \(\lambda_2 =0.05JK/ m\).}
\end{figure}

\begin{figure}
\begin{center}
       \includegraphics[width=0.7\textwidth]{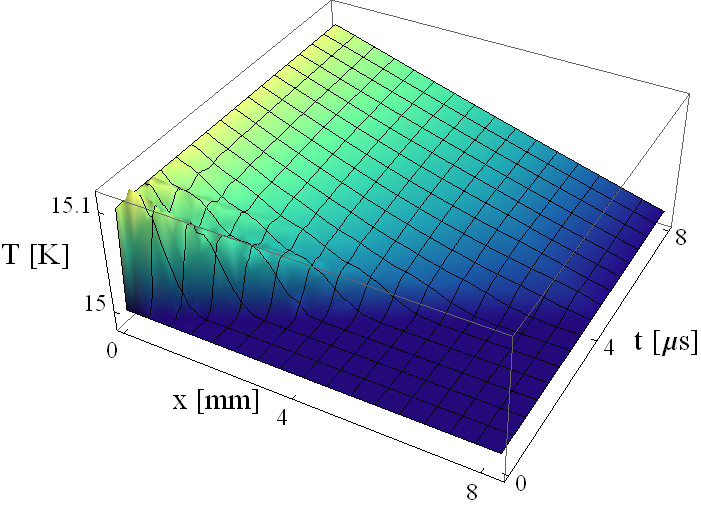}
\end{center}
\caption{ \label{GKfig}
Guyer-Krumhansl equation,  \(\tau  = 6.8 \mu s\),  \(k=10^{-12}ms^{3}/kgK\).}
\end{figure}

\begin{figure}
\begin{center}
       \includegraphics[width=0.7\textwidth]{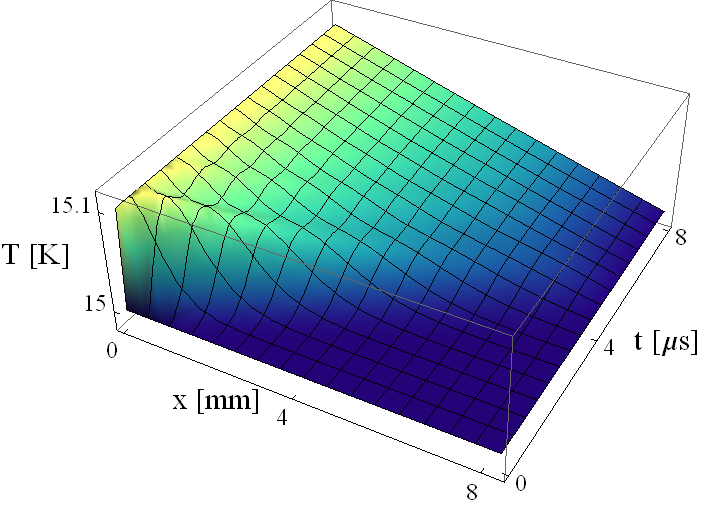}
\end{center}
\caption{ \label{JeffreysGKfig}
Jeffreys-Guyer-Krumhansl equation,  \(\tau = 6.8 \mu s,\ \lambda_2 =0.05JK/ m,\ k=10^{-12}ms^{3}/kgK\).}
\end{figure}

In the Figure \ref{MCVfig}  a solution of the Maxwell-Cattaneo equation is shown whith \(\tau= 0.68\mathrm{\mu s}\) and the other parameters are the same, \(\lambda_2\) and \(k\) are zero. The relaxation time is extracted from the measured \(v_{c} =1.95\mathrm{mm/\mu s}\). Here one may observe a ridge due to the finite propagation speed, and the damping of the ripples and of the reflected pulse due to the dissipative effect of heat conduction. Figures \ref{Jeffreysfig} and \ref{GKfig} show the solutions of the Jeffreys type and the Guyer-Krumhansl equations, where \(\tau\) is the same as in Figure \ref{MCVfig}, and \(\lambda_2 = 0.05\mathrm{JK/ m}\) (Jeffreys case) and \(k=10^{-12}\mathrm{ms^{3}/kgK}\) (Guyer-Krumhansl case) respectively. In both cases, the relaxation to the stationary solution is faster. In Figure \ref{JeffreysGKfig}, one can see a solution where none of the coefficients of equation \re{hceq} is zero . There the coefficients are \(\tau = 0.68\mathrm{\mu s}\), \(\lambda_2 =0.05\mathrm{JK/ m},\ k=10^{-12}\mathrm{ms^{3}/kgK}.\) In this case, one may observe the change of the speed of the signal propagation. 

\begin{figure}
\begin{center}
        \includegraphics[width=0.7\textwidth]{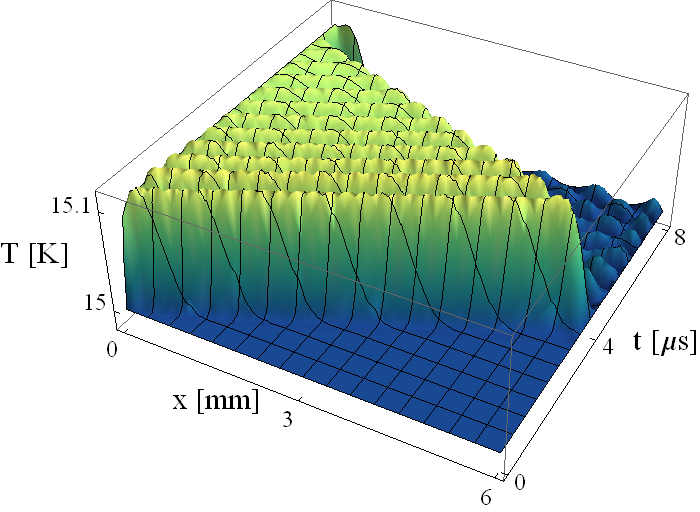}
\end{center}
\caption{ \label{GNfig}
Nondissipative Green-Naghdi type equation  \(\tau = 6.8 \mu s\).}
\end{figure}

\begin{figure}
\begin{center}
        \includegraphics[width=0.7\textwidth]{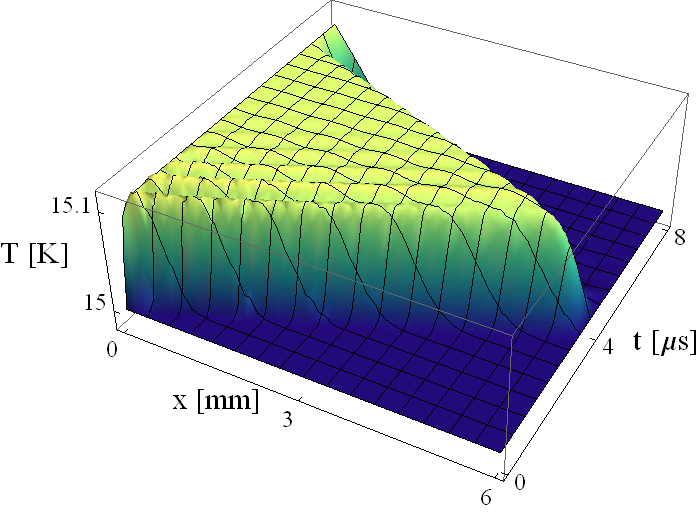}
\end{center}
\caption{ \label{GNJeffreysfig}
Green-Naghdi type equation with Jeffreys damping   \(\tau = 6.8 \mu s\), \( \lambda_2 =0.05JK/ m\).}
\end{figure}

\begin{figure}
\begin{center}
        \includegraphics[width=0.7\textwidth]{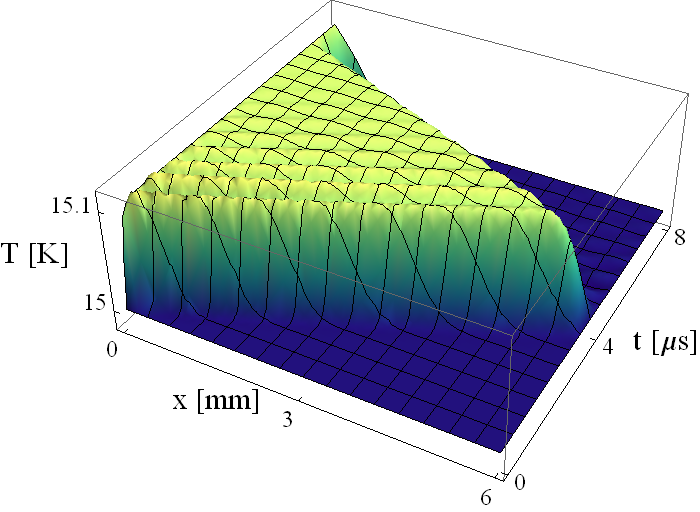}
\end{center}
\caption{ \label{GNGKfig}
Green-Naghdi type equation with Guyer-Krumhansl damping  \(\tau =  6.8 \mu s\),\ \(k=10^{-12}ms^{3}/kgK\).}
\end{figure}

\begin{figure}
\begin{center}
        \includegraphics[width=0.7\textwidth]{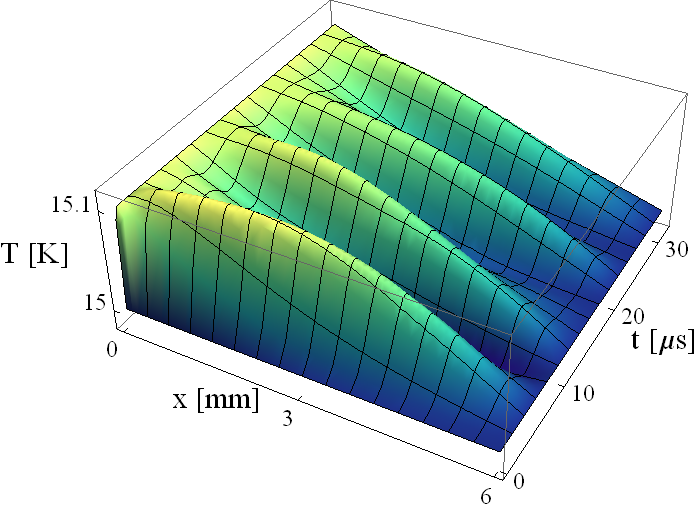}
\end{center}
\caption{ \label{dGNfig}
Green-Naghdi with both kinds of damping,  \(\tau = 6.8\mu s\), \(\lambda_2 = 0.1JK/ m,\ k=2\times10^{-11}ms^{3}/kgK \), longer duration.}
\end{figure}

The solution of the nondissipative GN type equation is given in Figure \ref{GNfig} with the same coefficient values as in case of MCV equation, \(\tau= 0.68\mathrm{\mu s,} \ \lambda_{1} =  4.61\times 10^6 \mathrm{WK/m}\). Figures \ref{GNJeffreysfig} and \ref{GNGKfig} show that Jeffreys or Guyer-Krumhansl dissipation do not damp the reflected travelling wave (parameter values:\ \(\tau = 0.68\mathrm{\mu s}\),  \(\lambda_2 = 0.05\mathrm{JK/ m,}\) respectively \(k=10^{-12}\mathrm{ms^{3}/kgK}\) in Figure \ref{GNGKfig}. Finally Figure \ref{dGNfig} shows that, for the general GN equation, the damping effects is due to the additional terms are different, from which can be observed for the GK-Jeffreys type one. Using the   \(\tau = 0.68\mathrm{\mu s}\), \(\ \lambda_2 = 0.1\mathrm{JK/ m},\ k=2\times10^{-11}\mathrm{ms^{3}/kgK}\) parameter values, one gets a qualitatively different damping than with the same parameter values for the heat conduction equation \re{gGNeq}.  


\section{Summary and conclusions}

We have introduced a vectorial internal variable and a current multiplier as a simple and general characterization of the deviation from local equilibrium of the entropy and the  entropy current density functions, respectively. Then we have derived the resulting connection between the heat flow and temperature, and obtained a common generalization of several known heat conduction equations. Consequently, the emerging heat conduction equation \re{geneq} is independent of the micro- or mesoscopic mechanism of heat conduction as long as the general conditions of this approach---namely, the second law and the existence of extra field quantities---are fulfilled. In this sense, the found heat conduction equation is \textit{universal}. We have demonstrated this kind of universality by the derivation of two common simple heat conduction mechanisms in our framework. 

The analysis of the different specific examples in the light of the thermodynamic constraints, stemming from the non-negative entropy production, revealed that: 
\begin{itemize}
\item If the main Fourier coefficient \((l_1)\) is not zero then MCV is always extended by the characteristic nonlocal term of the Jeffreys type equation.
\item Pure MCV, GK and GN type equations are related to a Casimir type cross effect between the internal variable and thermal parts of the entropy production.
\item The nondissipative wave equation of the GN type model  was derived and interpreted in the framework of  non-equilibrium thermodynamics.      
\item The strictly linear thermodynamic constitutive equations \re{o1}--\re{o3} with constant coefficients lead to a nonlinear evolution equation of the heat flow, where the nonlinearity cannot be absorbed in  special temperature dependent coefficients.
\item The general heat conduction equation \re{geneq} is remarkably stable in numerical calculations, whenever the inequalities \re{poz} are fulfilled. (For the link between stability and the second law of thermodynamic, see e.g. \cite{Mat05b})
\item With the natural choice of the internal variable as the non-Fourier part of  the heat flow,  a  physical interpretation of our internal variable was given. We found that two-temperature relaxation and heat flow splitting models can be interpreted in terms of the current multiplier. 
\end{itemize}

 It is worth mentioning that \re{Nyiriec} trivially corresponds to the usual generalisation {of the entropy current density}, when an additive constitutive term, the so-called M\"uller's \(K^i\) vector is introduced (see e.g. \cite{Mul85b}, p441).

{Another comparison that one can perform is related to the GENERIC approach in \cite{GrmEta11a}. Equations (42)-(49) in \cite{GrmEta11a} are in principle more general than ours from several points of view. On the other hand, the structure and derivation of the equations is rather different, therefore, the comparison is not straightforward. For example, the suggested evolution equations are balance like and the particular form of the entropy flux (first part of Eq. (46) in \cite{GrmEta11a}) looks structurally different from our simple generalization, preventing a direct comparison. } 

\section{Acknowledgements}
The authors are grateful to Arkadi Berezovski, J\"uri Engelbrecht,  Gyula Gr\'of, Bal\'azs Cz\'el and Joe Verh\'as for the discussions. The work was supported by the grant OTKA K81161 and  TÉT 10-1-2011-0061/ZA-15-2009.

\bibliographystyle{unsrt}

\end{document}